\documentstyle[12pt]{article}
\newcommand{\2}{\frac{1}{2}}

\begin{document}
\centerline{\Large\bf Bose-Einstein condensation in the }
\bigskip
\centerline{\Large\bf harmonic oscillator potential}
\vskip 10mm
\centerline{H. Haugerud and  F. Ravndal}
\centerline{\it Institute of Physics}
\centerline{\it University of Oslo}
\centerline{\it N-0316 Oslo, Norway}

\vspace{10mm}
{\bf Abstract:} {\footnotesize The Bose-Einstein condensation of a
dilute gas of rubidium-87 atoms
was achieved by cooling a small number of atoms in a magnetic trap. The
effective potential of the trap is to lowest order harmonic and under
these conditions we estimate the critical temperature and the heat
capacity of an ideal gas of Bose particles. Previously this
has been estimated within the semiclassical approximation.
However, because of the small number
of atoms, the resulting $T_c$ deviates slightly from our result obtained
by exact numerical calculations. For the gap of the heat capacity at the
transition point the deviation is $20\%$.}

\vspace{10mm}
In the remarkable experiment\cite{CORNELL} the rubidium vapor is
magnetically trapped and
evaporatively cooled below $170$ nK where a finite fraction of the
particles is observed to occupy the ground state. When
further lowering the temperature this fraction increases abruptly,
signaling a Bose-Einstein condensation.
The gas is very dilute and hence well approximated by an ideal gas.
The confining magnetic field is a
time-averaged, orbiting potential\cite{PETRICH} whose effective potential
to first order is a three-dimensional anisotropic harmonic oscillator.
The Bose-Einstein condensation of an ideal gas in such a potential has
been treated earlier in the semiclassical approximation\cite{KLEPPNER}.
We will first shortly review a part of this work using a slightly
different approach
leading to the same results.

The number of particles of an ideal Bose-gas is given by
\begin{equation}
N = N_0 + \sum_{i}\frac{1}{e^{(\varepsilon_i - \mu)/kT} - 1} \label{Nclasic}
\label{sum}
\end{equation}
where $N_0$ is the number of particles occupying the ground-state,
$\mu$ is the chemical potential and the sum runs over all one-particle
states. The semiclassical approximation consists of approximating
this sum by the phase-space integral
\begin{equation}
N = N_0 + \int \frac{d^3{\bf x} d^3{\bf p}}{(2\pi \hbar)^3}
	\frac{1}{e^{(H - \mu)/kT} - 1} \label{Nint}
\end{equation}
In our case the classical Hamiltonian is given by
$H = {\bf p}^2/2m + \2 m \omega^2(x^2 + y^2) + \2 m \omega_z^2$
and the integral can be evaluated at the transition point where $\mu = N_0 =
0$.
This gives the critical temperature
\begin{equation}
T_{c} = \frac{\hbar \omega}{k}
     \left(\frac{\omega_z}{\omega}\right)^{1/3}
     \left(\frac{N}{\zeta(3)}\right)^{1/3}	\label{T_{c}}
\end{equation}
where $\zeta(3) = 1.202$. Similarly the internal energy is given by
\begin{equation}
E(T) = \int \frac{d^3{\bf x} d^3{\bf p}}{(2\pi \hbar)^3}
	\frac{H}{e^{(H - \mu)/kT} - 1}
\end{equation}
The heat capacity $C = \partial E/ \partial T$ is
seen to be a function of $\partial \mu/ \partial T$. The latter can be
determined by taking the derivative of Eq.(\ref{Nint}) with
respect to the temperature\cite{KLEPPNER}. This enables us to calculate
the heat capacity below and above $T_{c}$:
\begin{eqnarray}
C(T_{c}^-) &=&  \frac{2\pi^4}{15\zeta(3)} Nk = 10.81 Nk \label{C1} \\
\Delta C = C(T_{c}^-) - C(T_{c}^+) &=&  \frac{54\zeta(3)}{\pi^2} Nk = 6.57 Nk
\label{C2}
\end{eqnarray}
The heat capacity is seen to exhibit a gap at the transition point, unlike
the ideal gas confined in a box.

In the experiment the number of particles was $N = 20000$ and the frequencies
of
the oscillators where given by $\omega_z/2\pi = 120 Hz$ and
$\omega = \omega_z/\sqrt{8}$. Eq.(\ref{T_{c}}) then gives  the transition
temperature
$T_{c} = 73.5$ nK. However, in a recent preprint\cite{BAYM} the frequency
is corrected to $\omega_z/2\pi = 208 Hz$, and will use this value in our
numerical estimates. This frequency gives a transition temperature $T_{c} =
127.4$
nK which is closer, but still not close to the observed transition temperature
at about $170$ nK.

To improve this method one may include interactions using the mean-field
approach, leading to corrections dependent of the scattering
length $a$\cite{KLEPPNER}, which for the rubidium gas is determined to be
in the range $85a_0 < a < 140a_0$, where $a_0$ is the Bohr
radius\cite{GARDNER}.
These corrections lowers the transition temperature to
$Tc = 125.6$ for $a = 110a_0$.

Another improvement is to go beyond semiclassical approximation which is
valid just for $kT \gg \hbar\omega$. In order to fulfill this requirement
at $T_c$ the number of particles must be quite large since
$kT_c/\hbar\omega \sim N^{1/3}$. However, when $\hbar\omega$ is comparable
to $kT$ in size, the exact sum of
Eq.(\ref{sum}) converges quite fast, and it is possible to evaluate the
sum over the quantized oscillator-states numerically.
Estimating the critical temperature in this way gives a $3\%$ correction
and $T_{c} = 123.4$ nK. This correction is seen to be twice as large as
the one due to interactions.

After determining $T_{c}$ the internal energy and specific heat may be
evaluated by replacing the phase-space integrals leading to
Eqs.(\ref{C1}-\ref{C2}) by a finite number of terms of the exact sums.
This gives
\begin{eqnarray}
C(T_{c}^-) &=& 10.20 Nk \\
\Delta C &=& 5.28 Nk
\end{eqnarray}
the latter differing $20\%$ from the semiclassical result, mainly due
to the shifted transition temperature.

The semiclassical approach thus turns out to be inaccurate when dealing
with such a small number of particles and instead the exact quantum sums
should be evaluated when calculating thermodynamic quantities. The
estimated transition temperature deviates from the observed. Probably
the dilute rubidium gas is well approximated by an ideal gas, at least
before the condensation takes place. So the crucial point is how well
the harmonic oscillator approximates the confining potential and if so,
how accurate the oscillator strengths may be determined.

\end{document}